\documentclass[aps,prd,reprint,superscriptaddress,nofootinbib,longbibliography]{revtex4-2}

\usepackage{amssymb,amsmath,bm,natbib}
\usepackage[dvipsnames]{xcolor}
\usepackage[colorlinks = true,
            linkcolor = Maroon,
            urlcolor  = Maroon,
            citecolor = Maroon]{hyperref}
\usepackage{microtype}
\usepackage{graphicx}
\usepackage{xspace}
\usepackage{bm}
\usepackage{bbm}
\usepackage{tikz}
\usetikzlibrary{decorations.markings}
\usetikzlibrary{decorations.pathmorphing}
\tikzset{snake it/.style={decorate, decoration=snake}}
\usetikzlibrary{matrix,calc}

\definecolor{harvardcrimson}{rgb}{0.79, 0.0, 0.09}

\begin{document}

\title{Hunting axion dark matter with antiferromagnets: A case study with nickel oxide}

\author{Pier~Giuseppe~Catinari}
\email{piergiuseppe.catinari@uniroma1.it}
\affiliation{Dipartimento di Fisica, Sapienza Universit\`a di Roma, Piazzale Aldo Moro 2, I-00185 Rome, Italy}
\affiliation{INFN Sezione di Roma, Piazzale Aldo Moro 2, I-00185 Rome, Italy}

\author{Angelo~Esposito}
\email{angelo.esposito@uniroma1.it}
\affiliation{Dipartimento di Fisica, Sapienza Universit\`a di Roma, Piazzale Aldo Moro 2, I-00185 Rome, Italy}
\affiliation{INFN Sezione di Roma, Piazzale Aldo Moro 2, I-00185 Rome, Italy}

\author{Shashin~Pavaskar}
\email{pavaskar@illinois.edu}
\affiliation{Department of Physics, University of Illinois at Urbana-Champaign, Urbana, Illinois 61801, USA}
\affiliation{IQUIST, University of Illinois at Urbana-Champaign, Urbana, Illinois 61801, USA}

\date{\today}

\begin{abstract}
\noindent We show that nickel oxide, which is already a very promising target to look for sub-MeV dark matter scattering, can be employed to hunt axion dark matter, with masses in the meV range and couplings to electrons allowing them to potentially be QCD axions. We describe the interactions between axions and the collective excitations of nickel oxide in terms of an effective field theory, built solely out of symmetry arguments. The processes of conversion into one or two excitations provide, respectively, a narrowband and a broadband channel for the axion search, and the possibility of varying an external magnetic field up to a phase transition point allows one to cover a large portion of a yet unexplored parameter space, reaching axion masses down to few fractions of an meV. Our results underline nickel oxide as an ideal candidate for a multipurpose target for light dark matter searches.
\end{abstract}

\maketitle


\section{Introduction}

\noindent Direct dark matter searches for both heavy Weakly Interacting Massive Particles~\cite[e.g.,][]{LUX:2016ggv,SuperCDMS:2018mne,XENON:2019gfn,CRESST:2019jnq,PandaX-II:2020oim} and axions~\cite[e.g.,][]{Barbieri:2016vwg,QUAX:2020adt,Salemi:2021gck,ADMX:2021nhd,Bartram:2024ovw,QUAX:2024fut,CAST:2024eil,Yan:2019dar,XENON:2022ltv,Capozzi:2020cbu,AxionLimits}, have so far reported null results. Concerning the axion, in particular, only a handful of experiments have (or will have) enough sensitivity to probe the QCD axion region, i.e. the right masses and couplings to potentially offer a solution to the strong CP~problem. In order to keep expanding the scope of our searches, and probe yet unexplored regions in parameter space, new ideas and detection strategies are needed.

A good deal of work has been recently devoted toward taking advantage of the rich spectrum of collective excitations featured by different condensed matter systems, trying to employ them to look for sub-MeV dark matter scattering and axion absorption, probing new values of their possible mass~\cite[e.g.,][]{Guo:2013dt,Hochberg:2016ajh,Schutz:2016tid,Knapen:2016cue,Hochberg:2016sqx,Knapen:2017ekk,Hertel:2018aal,Griffin:2018bjn,Marsh:2018dlj,Acanfora:2019con,Trickle:2019ovy,Caputo:2019cyg,Caputo:2019xum,Trickle:2019nya,Griffin:2019mvc,Baym:2020uos,Mitridate:2020kly,Chigusa:2020gfs,Caputo:2020sys,Matchev:2021fuw,You:2022pyn,Campbell-Deem:2022fqm,Esposito:2022bnu,Berlin:2023ubt,Chigusa:2023szl}. For a review, see~\cite{Kahn:2021ttr,Zurek:2024qfm}.

In this work we show that nickel oxide (NiO), a well studied antiferromagnet~\cite[e.g.,][]{hutchings1972measurement,milano2004effect,rezende2019introduction}, offers an optimal target to probe the electron coupling of axion dark matter with masses in the unexplored range $m_a \sim \mathcal{O}(0.1 - 10) \text{ meV}$. (Other proposals valid for similar masses are~\cite{Mitridate:2020kly,Chigusa:2020gfs,Berlin:2023ubt}.) It has already been shown that NiO provides an ideal target also to look for sub-MeV dark matter scattering from spin-dependent interactions~\cite{Esposito:2022bnu}. Among other reasons, this is because its vanishing magnetization ensures that the emission of multiple collective excitations (the so-called magnons) is allowed. This enables the transfer of a large fraction of the dark matter energy even at masses as low as the keV.

As we will argue, NiO is also very efficient at absorbing axions, with masses in the same range as the typical energies of its magnon modes. In particular, the process of axion conversion into a single excitation provides (at least) two resonant values for a narrowband search, corresponding to the two lowest lying magnon modes. By tuning an external magnetic field one can also make one of the magnons arbitrarily light, giving access to axion masses even below the meV range. The conversion into two magnons, instead, offers the possibility for a {\it broadband} channel: the system is able to absorb axions in a wide range of masses by simply adjusting the total energy and relative momentum of the two outgoing excitations. 

We work within an effective field theory (EFT) for the description of magnons and their interactions, valid at low energies and long wavelengths~\cite{chandra1990quantum,Burgess:1998ku,Pavaskar:2021pfo,Brauner:2024juy}, which we extend to incorporate the rich phenomenology of NiO~\cite{theory}. 
In this framework, magnons are the (pseudo-)Goldstone bosons associated with the breaking of non-relativistic spin symmetry, as induced by the antiferromagnetic ground state. The main advantage of this EFT is that it is completely dictated by symmetry, thus bypassing the need for a microscopic description of the system. 
The intricacies of the short distance physics are encoded in a few effective coefficients, which we match from data.
This has a twofold advantage. On the one hand, the EFT formulation allows one to clearly disentangle those phenomena that are universal, i.e., common to a broad class of systems, to those that are specific to a given short distance realization of these systems. On the other hand, it helps formulating the current problem in a language that is closer to high energy physics, thus allowing one to borrow concepts and techniques from this field.

\vspace{1em}

\noindent{\it Conventions:} We use natural units, $\hbar = c = 1$, and follow an index notation with $i,j,k = 1,2,3$ and $a,b=1,2$.


\section{The EFT}

\noindent We start by summarizing the EFT for magnons in NiO, which has been developed in full detail in~\cite{theory}. After that, we introduce the axion, and describe how to incorporate it into the above mentioned EFT. Throughout our treatment, we will assume to work at zero temperature, as motivated by the standard setup for direct dark matter searches.


\subsection{Magnons in NiO} \label{sec:EFT}

\noindent A magnetic material can be thought of as a collection of spins arising from the elementary spins of the unpaired electrons localized around each lattice site, with the exchange and superexchange interaction between these electrons providing an effective coupling between different spins. In an antiferromagnet, these couplings are such that, below a certain critical temperature, the ground state of the system develops a long-range order, with neighboring spins antiparallel to each other and all aligned along one direction. We take this to be the $z$-axis. For NiO, this happens below the temperature $T_{\rm N} \simeq 523 \text{ K}$, called the N\'eel temperature~\cite[e.g.,][]{tomlinson1955high,hutchings1972measurement}.

\begin{table*}[t]
    \centering
    \begin{tabular}{c c c c c c c}
        \hline\hline
        $v_\theta$ & $c_1 \; [\,{\rm MeV/
        \AA}\,]$ & $\mu/\mu_{\rm B}$ & $\lambda_x \; [ \, {\rm meV}^2 \,]$  & $\lambda_z \; [ \, {\rm meV}^2 \,]$ & $\Lambda_{\rm UV} \; [\,{\rm keV}\,]$ & $\rho_{\rm T} \; \big[\,{\rm g/cm}^3 \,\big]$ \\\hline
        $1.3 \times 10^{-4}$ & 0.58 & 2.18 & 9.80 & 0.17 & 0.6 & 6.7 \\\hline\hline
    \end{tabular}
    \caption{Parameters for the EFT of magnons in NiO~\cite{theory}. $\mu_{\rm B}$ is the Bohr magneton, and the mass density, $\rho_{\rm T}$ is taken from~\cite{kannan2020nanostructured}. $\Lambda_{\rm UV}$ is estimated from~\cite{hutchings1972measurement} as the momentum for which the dispersion relation deviates from linear by 10\%. Note that while the EFT is general for any antiferromagnet with the same symmetry breaking pattern, the value of the parameters crucially depends on the microscopic structure of the order parameter and of the crystal.
    }
    \label{tab:coeff}
\end{table*}

In this context, one can define an order parameter, the so-called N\'eel vector, consisting of a staggered sum of all the spins in the material, i.e., $\bm{\mathcal{N}} \equiv \sum_i (-1)^i \bm{\mathcal{S}}_i$, where $\bm{\mathcal{S}}_i$ is the spin of the $i$-th lattice site and $(-1)^i$ is positive for spins on even sites, and negative for those on odd sites. In the ordered ground state, the N\'eel vector acquires a non-zero expectation value, $\langle \bm{\mathcal{N}} \rangle \propto \hat{\bm z}$. From the symmetry viewpoint, this spontaneously breaks the internal spin rotations to the subgroup of rotations that leave the order parameters unchanged, i.e. those around the $z$-axis. The symmetry breaking pattern is then ${\rm SO}(3) \to {\rm SO}(2)$, and magnons are nothing but the associated Goldstone bosons. As such, they can be described by an EFT.

A convenient way of parametrizing them is in terms of local modulations of the order parameter, $\hat{\bm n} = e^{i \theta^a \, S_a} \!\cdot\! \hat{\bm z}$, where $S_a = \{ S_1, S_2 \}$ are the broken SO(3) generators, with $(S_i)_{jk} = -i \epsilon_{ijk}$, and $\theta^a(x)$ are the magnon fields. The most general, lowest order Lagrangian describing magnons at low energies and small momenta is simply given by~\cite[e.g.,][]{Burgess:1998ku}
\begin{align} \label{eq:Ltheta}
    \mathcal{L}_\theta ={}& \frac{c_1}{2} \left[ \left( \partial_t \hat{\bm n} \right)^2 - v_\theta^2 \left( \nabla_i \hat{\bm n} \right)^2 \right] \,.
\end{align}
Here $v_\theta$ is the magnon propagation speed, and it can be determined from their dispersion relation, obtaining $v_\theta \simeq 1.3 \times 10^{-4}$~\cite{Esposito:2022bnu}. 
The coefficient $c_1$ can instead be matched to the parameters of the microscopic spin Hamiltonian using neutron scattering data~\cite{Esposito:2022bnu}. We report the values of all the effective coefficients in Table~\ref{tab:coeff}.

The phenomenology of NiO is, however, richer than this. Specifically, its magnon modes actually present a small but non-zero gap. This is due to tiny explicit breaking effects, turning the magnons into {\it pseudo}-Goldstones. 
To the best of our knowledge, the microscopic mechanism generating these gaps has not been completely understood. Data from Raman and Brillouin scattering experiments seem to point to magneto-crystalline anisotropies and dipolar interactions between spins as the most plausible explanation~\cite{milano2004effect}. 
Yet, the simplest model which correctly reproduces the spectrum of the two lowest lying magnon modes only includes two anisotropic terms with no dipole interactions, as done, for example, in~\cite{hutchings1972measurement,rezende2019introduction}. These anisotropies explicitly break the internal SO(3), and are encoded in the EFT by two additional operators~\cite{theory},
\begin{align}
    \mathcal{L}_{\rm anis} = c_1 \left[ \, \lambda_z \, \hat{n}_z^2 - \lambda_x \, \hat{n}_x^2 \, \right] \,,
\end{align}
where $\lambda_z,\lambda_x >0$.  The anisotropic couplings can also be matched to the parameters of the microscopic Hamiltonian~\cite{theory}, and their values are again in Table~\ref{tab:coeff}. The anisotropies in the $x$- and $z$-directions are often dubbed ``hard'' and ``easy'' axis anisotropies, respectively~\cite{rezende2019introduction}.

In order to scan different axion masses, one needs to vary the magnon gap. This can be achieved by placing the NiO sample in a constant magnetic field, which we take to be aligned with the easy axis, $\text{\bf H} = \text{H} \, \hat{\bm z}$.
This can be accounted for in the EFT by promoting the ordinary time derivative to a covariant one in the Lagrangian~\eqref{eq:Ltheta}, $\partial_t \hat{\bm n} \to \partial_t \hat{\bm n} + \mu \text{\bf H} \times \hat{\bm n}$, such that the Lagrangian is now invariant under simultaneous gauge transformations of $\hat{\bm n}$ and $\text{\bf H}$, with the latter playing the role of a vector potential.
Indeed, at the level of the Hamiltonian, this procedure returns the standard Zeeman coupling~\cite{leutwyler1994nonrelativistic,Brauner:2024juy,theory}. Here $\mu$ is the gyromagnetic ratio, again given in Table~\ref{tab:coeff}. 

All together, the EFT for magnons in NiO is given by
\begin{align} \label{eq:LEFT}
    \begin{split}
        \mathcal{L}_{\rm EFT} ={}& \frac{c_1}{2} \left[ \big( \partial_t \hat{\bm n} + \mu \text{\bf H} \times \hat{\bm n} \big)^2 - v_\theta^2 \big( \nabla_i \hat{\bm n} \big)^2  \right] + \mathcal{L}_{\rm anis} \\
        ={}& \frac{c_1}{2} \bigg[ \big( \dot{\theta}^a - \mu \text{H} \epsilon^{ab} \theta^b \big)^2 - v_\theta^2 \big( \bm \nabla \theta^a \big)^2 \\
        & \quad\, - 2\lambda_z \big(\theta^a\big)^{2} - 2 \lambda_x \delta^{a2} \delta^{b2} \theta^a\theta^b + \mathcal{O}\big(\theta^4\big)  \bigg] \,,
    \end{split}
\end{align}
where we expanded in fluctuations around equilibrium.

From the solutions to the linearized equations of motion one finds that there are, as expected, two physical magnon modes, with dispersion relations given by~\cite{theory},
\begin{align} \label{eq:dispersion}
    \begin{split}
        \omega^2_{q,\pm} ={}& \mu^2 \text{H}^2 + \lambda_x + 2 \lambda_z + v_\theta^2 q^2 \\
        &\pm \sqrt{4 \mu^2 \text{H}^2 (\lambda_x + 2 \lambda_z + v_\theta^2 q^2) + \lambda_x^2} \,.
    \end{split}
\end{align}

The quantization of this theory is complicated by the presence of a one-derivative kinetic term, which does not allow one to diagonalize the quadratic Lagrangian~\eqref{eq:LEFT} by a local field redefinition.
The details of the quantization procedure have been developed in~\cite{theory}. 

Importantly, as the magnetic field increases, the gaps of the $+$ and $-$ modes in Eq.~\eqref{eq:dispersion} respectively increase and decrease. This feature can be used to scan over different axion masses. At $\mu_0\text{H} = \mu_0 \sqrt{2 \lambda_z}/\mu \simeq 4.62$~T,\footnote{In the presence of a material, the magnetic field $\text{H}$ is typically reported in oersted (Oe). In this work, we use $\mu_0 = 10^{-4}$~T/Oe to convert to Tesla, which is more common in the high energy literature.} the lightest modes becomes gapless and, for larger fields, the system undergoes a (first order) ``spin-flop'' phase transition to a different ground state, characterized by $\langle \hat{\bm n} \rangle = \hat{\bm y}$~\cite[e.g.,][]{rezende2019introduction,theory}. 
In this work, we are not interested in this regime.
However, we notice that thermal and radiative corrections are expected to change both the magnon dispersion relation, as well as the value of the critical magnetic field, but not too dramatically~\cite{machado2017spin}. They are also expected to induce a small gap even at the phase transition point.

It is also interesting to note that in the absence of magnetic fields, the magnon gaps are $\omega_{0,+/-}\sim \sqrt{\lambda_{x/z}}$, corresponding to $\mathcal{O}(\rm meV)$ energies. This is in contrast to ferrimagnets~\cite{Mitridate:2020kly}, where the gaps are $\mathcal{O}(\mu{\rm eV})$, and hence are more suitable to probe axion masses in that range.

Finally, our EFT breaks down for momenta larger than a strong coupling scale, $\Lambda_{\rm UV}$, when the details of microscopic system become relevant. This cutoff can be estimated, for example, as the
momentum for which the dispersion relation sensibly deviates from linearity, indicating that higher derivative
terms in the effective Lagrangian become relevant. In the rest of this work, we always ensure that momenta and energies are below, respectively, $\Lambda_{\rm UV}$ and $v_\theta \Lambda_{\rm UV}$, the latter being the typical microscopic energy scale.


\subsection{Axion-magnon coupling}

\noindent We are interested in the axion-electron coupling, which is ultimately responsible for the interaction between axions and magnons. The corresponding Lagrangian is
\begin{align}  
    \begin{split}
        \mathcal{L}_a = \frac{g_{aee}}{2m_e} \partial_\mu a \, \bar e \gamma^\mu \gamma^5 e &\xrightarrow{{\rm n.r.}}{} - \frac{g_{aee}}{m_e} \bm \nabla a \cdot \left( e_{\rm nr}^\dagger \frac{\bm \sigma}{2} e_{\rm nr} \right)
        \\
        &\xrightarrow{\,{\rm IR}\;}{} - \frac{g_{aee}}{m_e} \bm \nabla a \cdot \bm s
    \end{split}
\end{align}
where we first took the non-relativistic limit as described in~\cite{Esposito:2022bnu} (see also~\cite{Krnjaic:2024bdd}), with $e_{\rm nr}$ the non-relativistic electron field, neglecting terms suppressed by the ratio of the axion mass to the electron mass. 
In going to the second line, we used the fact that $e_{\rm nr}^\dagger \bm \sigma e_{\rm nr}/2$ is the electron spin density. At low energies, it must be expressed in the terms of the relevant degrees of freedom, which in this case are the magnons. Note that we have neglected an additional term, which scales as the electron velocity. For most materials, this is subleading with respect to the coupling included here. For a more detailed discussion we refer the reader to~\cite{Wang:2021dfj,Berlin:2023ubt}. The inclusion of this additional effect is beyond the scope of the present work.

Now, the spin density is nothing but the Noether current associated with the original SO(3) symmetry which, given the Lagrangian~\eqref{eq:LEFT}, reads as
\begin{align} \label{eq:s}
    s^i ={}& c_1 \! \left[\big( \partial_t \hat{\bm n} \times \hat{\bm n} \big)^i + \mu (\text{\bf H} \cdot \hat{\bm n}) \, \hat n^i \right] \\
    \simeq{}& c_1 \!\left[ \delta^{ia} \big( \dot\theta^a - \mu \text{H} \epsilon^{ab} \theta^b \big) - \delta^{i3} \theta^a \big( \epsilon^{ab} \dot\theta^b + \mu \text{H} \, \theta^a \big)\right] \,, \notag
\end{align}
where we omitted constant terms and neglected higher order contributions in the small fluctuations. 

With this at hand, and given the expression for the canonical fields reported in~\cite{theory}, one easily finds the amplitudes, $i\mathcal{M}$, for the axion-magnon conversions:
\begin{subequations}
\begin{align}
    \begin{tikzpicture}
        [decoration={markings, mark = at position 0.6 with {\arrow{stealth}}}, baseline = -0.25em]
        \draw [postaction={decorate},thick] (0,0) -- (0.7,0); 
        \draw [postaction={decorate},thick,dashed] (0.7,0) -- (1.4,0); 
        \node at (0.7,0) [circle,fill,inner sep=1.1pt]{};
        \node at (0.15,0.2) {\footnotesize $\bm p$};
        \node at (1.15,0.22) {\footnotesize $\bm q, \alpha$};
    \end{tikzpicture} \!={}& i  \frac{g_{aee}\sqrt{c_1}}{m_e}  p_a  \! \left[ i \omega_{q,\alpha} \mathcal{Z}^a_{q,\alpha} \! +\! \mu \text{H} \epsilon^{ab}  \mathcal{Z}^b_{q,\alpha}  \right]^* \!, \\
    \begin{split}
    \begin{tikzpicture}
        [decoration={markings, mark = at position 0.6 with {\arrow{stealth}}}, baseline = -0.25em]
        \draw [postaction={decorate},thick] (0,0) -- (0.7,0); 
        \draw [postaction={decorate},thick,dashed] (0.7,0) -- (1.4,0.6);
        \draw [postaction={decorate},thick,dashed] (0.7,0) -- (1.4,-0.6);
        \node at (0.7,0) [circle,fill,inner sep=1.1pt]{};
        \node at (0.15,0.2) {\footnotesize $\bm p$};
        \node[rotate=34] at (1.10,0.6) {\footnotesize $\bm q, \alpha$};
        \node[rotate=-34] at (1.10,-0.65) {\footnotesize $\bm k, \beta$};
    \end{tikzpicture} \!={}& i \, \frac{g_{aee}}{m_e} \, p_z \big[ i \left( \omega_{q,\alpha} - \omega_{k,\beta} \right) \epsilon^{ab} \mathcal{Z}^a_{q,\alpha} \mathcal{Z}^b_{k,\beta} \\[-1.5em]
    & + 2\mu \text{H} \mathcal{Z}^a_{q,\alpha} \mathcal{Z}^b_{k,\beta} \big]^*, \label{eq:M2}
    \end{split}
\end{align}
\end{subequations}
where $\alpha,\beta = \pm$ run over the physical magnon states, with dispersion relations as in Eq.~\eqref{eq:dispersion}. The coefficients $\mathcal{Z}^a_{q,\alpha} \equiv \langle 0 | \theta^a(0) | \bm q, \alpha \rangle$ are the overlap functions between the magnon fields and the physical magnon states. They are given by~\cite{theory}
\begin{subequations}
    \begin{align}
        {|\mathcal{Z}^1_{q,\pm}|} ={}& \sqrt{\frac{1}{2} \pm \frac{2\mu^2 \text{H}^2 - \lambda_x}{2\sqrt{4\mu^2 \text{H}^2 (v_\theta^2 q^2 + \lambda_x + 2\lambda_z) + \lambda_x^2}}} \,, \\
        \mathcal{Z}_{q,\pm}^2 ={}& i \left(\frac{v_\theta^2 q^2 - \mu^2 \text{H}^2 + 2\lambda_z - \omega_{q,\pm}^2}{2\mu \text{H} \,\omega_{q,\pm}} \right) \mathcal{Z}_{q,\pm}^1 \,,
    \end{align}
\end{subequations}
where the common overall phase is arbitrary. 


\section{Event rates}

\noindent We can now compute the axion absorption event rates per unit target mass. For a target material of density $\rho_{\rm T}$ (see Table~\ref{tab:coeff}), this is given by
\begin{align} \label{eq:R}
    R({\hat{\bm v}}_{\rm e})= \frac{\rho_a}{\rho_{\rm T} m_a} \int d^3 v \, f(|\bm v + \bm v_{\rm e}|) \, \Gamma(\bm v) \,.
\end{align}
Here $f(v)$ is the dark matter velocity distribution in the Milky Way. To compare with other proposals, we take this as a truncated Maxwellian as in the standard halo model, with dispersion $v_0 = 230$~km/s, escape velocity $v_{\rm esc} = 600$~km/s and boosted
with respect to the Galactic rest frame by the Earth velocity, $v_{\rm e} = 240$~km/s~\cite{Piffl:2013mla,monari2018escape}. We also take the axion dark matter density to be $\rho_a = 0.4$~GeV/cm$^3$~\cite{Piffl:2013mla,ou2024dark}.

One of the advantages of employing an EFT approach like we do, and thus formulating the problem in terms of local Lagrangians, lies in the fact that the rate $\Gamma$ can now be computed in the standard way, as explained in any particle physics textbook.


\subsection{One-magnon absorption and directionality}

\noindent For the conversion into a single magnon, momentum conservation enforces its momentum to be equal to the axion one, $\bm q = m_a \bm v$. Since we will be interested in masses $m_a \lesssim 1$~eV, this momentum is always extremely small, and can be set to zero everywhere without any appreciable corrections, as one can check explicitly. In light of this, the absorption rate is
\begin{align} \label{eq:Gamma_alpha}
    \Gamma_\alpha ={}& \frac{\pi m_a |g_{aee}|^2}{2 m_e^2} \frac{c_1}{\omega_{0,\alpha}} \big|v^a V_\alpha^a\big|^2  \, \delta(m_a - \omega_{0,\alpha}) \\
    \to{}& \frac{m_a |g_{aee}|^2}{ m_e^2} \frac{c_1}{\omega_{0,\alpha}} \big|v^a V_\alpha^a\big|^2 \frac{m_a \omega_{0,\alpha} \gamma_\alpha}{\big( m_a^2 - \omega_{0,\alpha}^2 \big)^2 + \big( \omega_{0,\alpha} \gamma_\alpha \big)^2} \,, \notag
\end{align}
where $V_\alpha^a \equiv i \omega_{0,\alpha} \mathcal Z_{0,\alpha}^a + \mu \text{H} \epsilon^{ab} \mathcal{Z}_{0,\alpha}^b$, and we regulated the energy $\delta$-function with a Breit-Wigner distribution, accounting for the magnon finite width, $\gamma_\alpha$. This width is roughly $\gamma_\alpha = \mathcal{O}(1 - 100) \text{ } \mu{\rm eV}$~\cite{bayrakci2006spin,bayrakci2013lifetimes}, and it depends on various properties of the material, such as Umklapp scattering, temperature, and boundary effects~\cite{bayrakci2013lifetimes}. To the best of our knowledge, this information is not available for NiO. We thus follow the conservative approach of~\cite{Mitridate:2020kly} and treat $\gamma_\alpha/\omega=\gamma_\alpha/m_a$ as a free parameter.

The event rate has a non-trivial dependence on the direction of the Earth velocity. This is because antiferromagnets feature a preferential direction, as set by the order parameter. This directionality can be exploited as an efficient strategy to discriminate between signal and background~\cite[e.g.,][]{Spergel:1987kx,Mayet:2016zxu}. As we show, in NiO this feature is quite remarkable, with a modulation following the direction of the Earth velocity of about 300\%.

\begin{figure}
    \centering
    \includegraphics[width=\columnwidth]{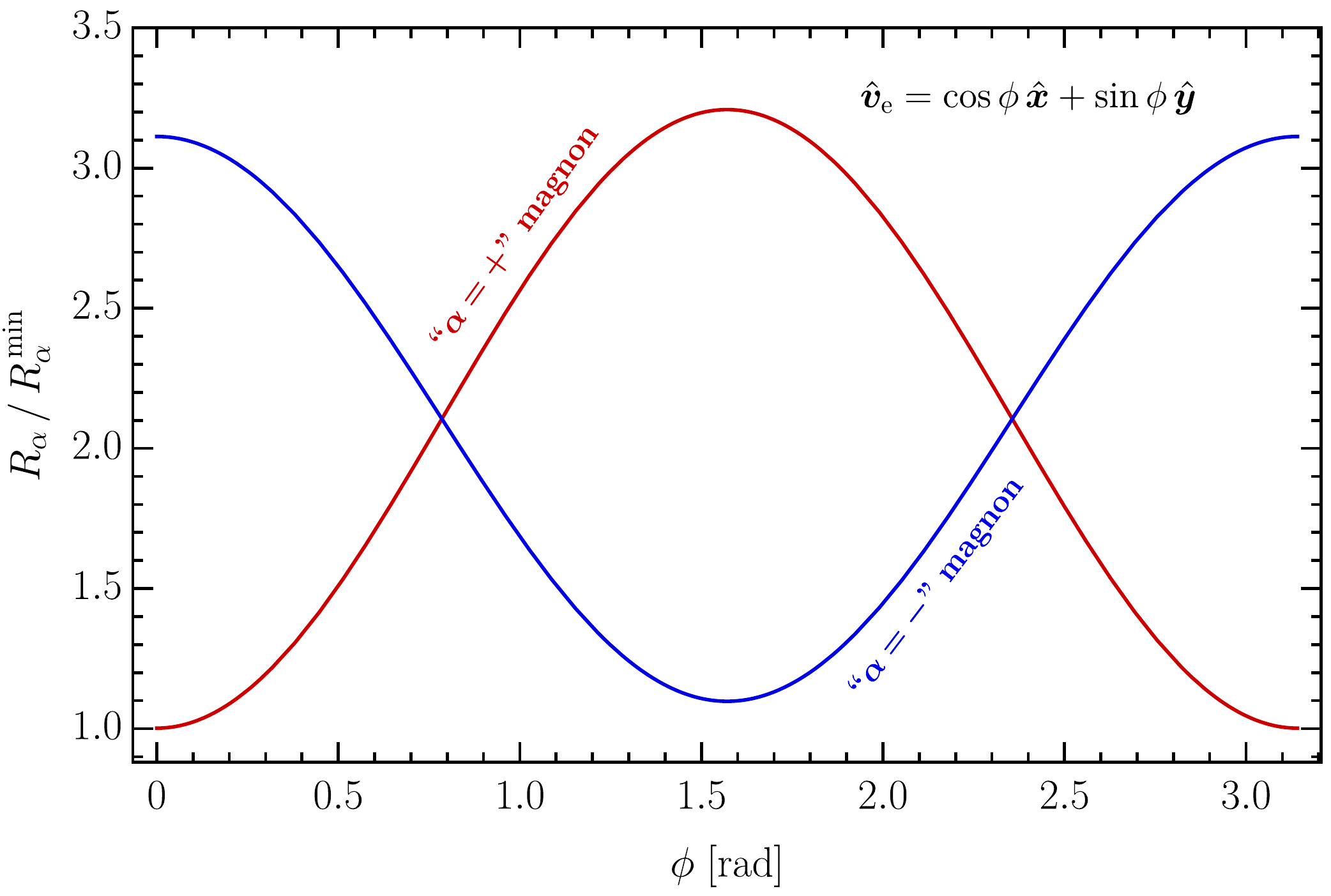}
    \caption{One-magnon event rate as a function of the direction of the Earth velocity, taken to be on the $xy$-plane, i.e. $\hat{\bm v}_{\rm e} = \cos \phi \, \hat{\bm x} + \sin\phi \, \hat{\bm y}$, at $\mu_0 \text{H}=1$~T. The rate is normalized to its minimum value, achieved for $\hat{\bm v}_{\rm e} = \hat{\bm z}$. Note that the minima of the curve lie very close to $R_\alpha/R_\alpha^{\rm min} = 1$. This means, for example, that the $R_+$ rates evaluated at $\bm v_{\rm e} = \hat{\bm z}$ and $\bm v_{\rm e} = \hat{\bm x}$ are very similar. This is because the $V_\alpha^a$ vectors have one component much larger than the other, as a consequence of the large hierarchy between the effects due to the external magnetic field, and those due to the intrinsic anisotropies.}
    \label{fig:modulation}
\end{figure}

After averaging Eq.~\eqref{eq:Gamma_alpha} over the dark matter velocity distribution, the rate depends on the Earth velocity as
\begin{align}
    R_\alpha(\hat{\bm v}_{\rm e})  \propto \frac{\langle \bm v_\perp^2 \rangle}{2} V_\alpha^a (V_\alpha^a)^* + \big| v_{\rm e}^a V_\alpha^a \big|^2 \,,
\end{align}
where $\bm v_\perp$ is a velocity perpendicular to the order parameter (i.e., on the $xy$-plane), and directionality is due to the second term. If the axion has the right mass to emit an $\alpha = +$ magnon, the rate is maximum when $\hat{\bm v}_{\rm e} = \hat{\bm y}$, while for $\alpha = -$, this happens when $\hat{\bm v}_{\rm e} = \hat{\bm x}$. In Fig.~\ref{fig:modulation} we report the directional dependence of the event rate. As anticipated, the signal modulation is very strong. However, we point out that realistic materials present various domains, each with a different orientation of the order parameter. This effect could decrease the modulation.\footnote{We are grateful to G.~Marocco for pointing this out.}

\begin{figure}
    \centering
    \includegraphics[width=\columnwidth]{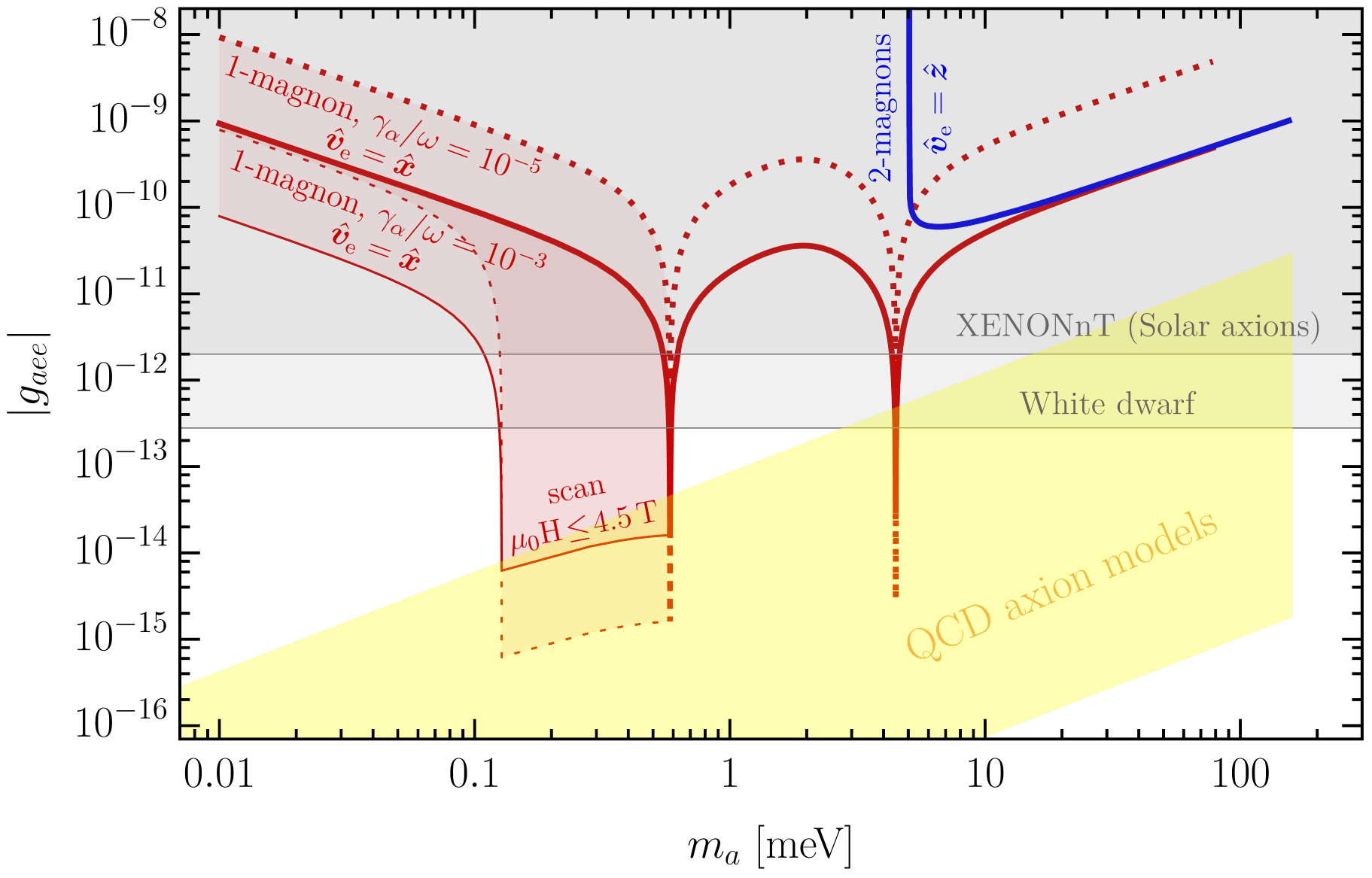}
    \caption{Projected reach at 95\% CL for a kilogram of material and a year of exposure, assuming no background. For single magnon absorption, we treat $\gamma_\alpha/\omega$ as a free parameter. The one- and two-magnon channels have been computed, respectively, for $\hat{\bm v}_{\rm e} = \hat{\bm x}$ and $\hat{\bm v}_{\rm e} = \hat{\bm z}$. The red shaded regions are obtained for a magnetic field varying in the range $\mu_0 \text{H}\in[0,4.5]$~T. The same scan for the higher magnon mode is not visible in logarithmic scale. The projections are interrupted when the energy/momenta involved exceed the EFT cutoff. The yellow band corresponds to the QCD axion models reviewed, for example, in~\cite{Irastorza:2018dyq,DiLuzio:2020wdo}, and is stopped in correspondence to the cosmological bound determined in~\cite{Bianchini:2023ubu}. The white dwarf bound is taken from~\cite{MillerBertolami:2014rka}, and the XENONnT one from~\cite{XENON:2022ltv} (see also~\cite{Caputo:2024oqc}).
    }
    \label{fig:projection}
\end{figure}

In Fig.~\ref{fig:projection} we instead show the projected reach
obtained assuming three events per year for a kilogram of NiO, in the ideal absence of background. The total rate is given by $R = R_+ + R_-$, and we evaluate it in the $\hat{\bm v}_{\rm e} = \hat{\bm x}$ configuration. The result obtained for the $\hat{\bm v}_{\rm e} = \hat{\bm y}$ is analogous, differing by no more than order 1. We show results for both $\gamma_\alpha / \omega = 10^{-3}$ and $10^{-5}$, with the second one being rather optimistic and likely achievable only in very pure samples. We also scan different values of the external magnetic field, in the range $\mu_0 \text{H}\in [0,4.5]$~T. Crucially, when this approaches the critical value (see Sec.~\ref{sec:EFT}), the one-magnon process probes lower and lower values of the axion mass, as again shown in Fig.~\ref{fig:projection}. This makes the $H$-field scanning in NiO potentially a very powerful avenue to cover a large portion of the parameter space.

\subsection{Two-magnon absorption}

\noindent As explained also in~\cite{Esposito:2022bnu}, antiferromagnets offer the possibility of multi-magnon events, which are instead strongly suppressed by symmetry in ferrimagnets~\cite{Trickle:2019ovy}. In the context of axion absorption, these allow for a broadband channel, since the system can absorb axions with different masses by simply changing the energy and relative momentum of the multiple excitations, without the need for any external tuning. In our EFT language, the matrix element for two-magnon absorption is in Eq.~\eqref{eq:M2}.

In this case the magnons are effectively emitted back-to-back, $\bm p \simeq - \bm k$, since again the axion momentum is irrelevant.  This considerably simplifies the calculation of the event rate, which is now the sum over all possible outgoing magnons $R = R_{++} + R_{+-} + R_{-+} + R_{--}$. 
Moreover, contrary to the one-magnon case, the maximum rate in this case is obtained for $\hat{\bm v}_{\rm e} = \hat{\bm z}$. 
The resulting projections are again reported in Fig.~\ref{fig:projection}. As one can see, the absorption into two magnons provides a broadband channel, which can be dominant when the magnons are particularly long-lived.


\section{Outlook}

\noindent NiO can be employed as an optimal target to probe the electron coupling of axion dark matter with masses around the meV and couplings down to the QCD axion ones, a region still experimentally uncharted. The one-magnon channel features a strongly directional signal, which can be key to an efficient background rejection strategy. Moreover, by varying the external magnetic field toward its critical ``spin-flop'' value, it is possible to probe axion masses that are, in principle, arbitrarily small, thus covering a large portion of the unexplored parameter space. Finally, the excitation of multiple magnons in the final state provides a broadband channel for axion absorption, which is the dominant signature in samples that allow for an especially narrow magnon mode.

Our results, together with the previous proposal to employ NiO as a target for sub-MeV dark matter scattering from spin-dependent interactions~\cite{Esposito:2022bnu}, make a strong physics case for the employment of this material as a {\it multi-purpose} target for light dark matter searchers, covering a brand new region of parameter space.

This also paves the way to a number of further investigations. For instance, our study does not deal with magnetic fields that are arbitrarily close to the critical point, as other effects might become relevant when the magnon becomes sufficiently light. It is also important to understand how the possible directional signal is affected by the presence of multiple domains in a realistic material, or if other material might perform better than NiO (see, e.g.,~\cite{Marocco:2025eqw}). In this second instance, one will have to work out another EFT, if useful.  Moreover, it is known that NiO features a number of additional light excitations~\cite{milano2004effect}, and that the magnon gaps can also be varied (quite a lot) by applying an external strain to the sample~\cite[e.g.,][]{kim2022giant}. Neither of these features are currently included in our EFT, and their implications for light dark matter searches have not been explored. In particular, if not suppressed by possible selection rules, the other light magnon modes could provide additional channels for axion absorption, which would allow one to cover an even larger region in parameter space. Finally, to turn this into a concrete proposal, one must address the question of what are the observable signatures associated with the emission of magnons in the NiO sample. This is also relevant in light of the fact that, once a concrete detection strategy is in place, the final excluded region could be different from the ideal one presented here. We leave these and other interesting directions for future work.


\begin{acknowledgments}
\noindent We are grateful to Giovanni~Carugno, Luca~Di~Luzio, Pedro~Guillaumon, Giacomo~Marocco and Francesca~Pucci for enlightening discussions. We are also particularly thankful to Andrea~Caputo for collaboration in the early stages of this project, and to Mauro~Valli for detailed and constructive comments on the manuscript. S.P. is supported in part by DOE Grant No. DE-SC0015655.
\end{acknowledgments}

\bibliography{biblio.bib}

\end{document}